# Topological heavy-tailed networks

Sunkyu Yu[1†], Xianji Piao[2§], and Namkyoo Park[3*]

[1]Intelligent Wave Systems Laboratory, Department of Electrical and Computer Engineering, Seoul National University, Seoul 08826, Korea

[2]Wave Engineering Laboratory, School of Electrical and Computer Engineering, University of Seoul, Seoul 02504, Korea

[3]Photonic Systems Laboratory, Department of Electrical and Computer Engineering, Seoul National University, Seoul 08826, Korea Seoul 08826, Korea

E-mail address for correspondence: sunkyu.yu@snu.ac.kr; piao@uos.ac.kr; nkpark@snu.ac.kr

**Abstract**

Although two-dimensional periodic structures have functioned as the primary platform for exploring topological phenomena, recent advances have substantially expanded this research boundary to include more intricate, aperiodic structures: quasicrystals, fractals, non-Euclidean lattices, and disordered materials. A network-based perspective not only offers a unified framework for classifying these diverse platforms based on their network connectivity but also unveils unexplored regimes of topological phenomena in complex networks. Here, we implement topological heavy-tailed networks, as an example of high-degree complex networks exhibiting topological phases. By developing a tight-binding model for the Apollonian network and a deterministic algorithm to assign nontrivial gauge fields to this aperiodic geometry, we compute the magnetic-flux-dependent energy spectrum—the Apollonian butterfly. Using spectral localizers, we characterize the topological features of the Apollonian butterfly, whose sensitivity is governed by lower-degree nodes—analogous to the controllability of complex networks. Our framework bridges topological physics and network science, introducing a connectivity-driven paradigm for the control of topological waves.

# Introduction

Two-dimensional (2D) crystalline structures have served as a foundational platform for exploring topological physics[1,2], owing to their closed-manifold Brillouin zones and bulk bandgaps, which enable well-defined topological invariants and edge states with practical relevance. A representative example is the Hofstadter model[3], in which a uniform magnetic flux penetrates each plaquette of a 2D periodic lattice, leading to magnetic subbands and topological gaps characterized by quantized Chern numbers—the celebrated Hofstadter butterfly. Over the past decade, this paradigm has been generalized far beyond 2D periodic lattices, enabling Chern topological phases even in the absence of discrete translational symmetry—for example, in quasicrystals[4-6], hyperuniform[7] or amorphous[8] media, lattices with dislocation[9] and disclination[10] defects, and non-Euclidean lattices[11]—while retaining gapped spectra.

Within the tight-binding picture, the 2D topological systems listed above can be modelled as graph networks with magnetic features[12-14], where each atomic state is represented by a node and each interatomic coupling by a complex-valued link associated with a nontrivial gauge. This network perspective enables the classification of topological platforms in terms of their network geometry. For example, the conventional Hofstadter lattice corresponds to the {4,4} regular magnetic graph, where the Schläfli symbol {$p,q$} denotes a regular tiling by $p$-gons with $q$ polygons meeting at each vertex. Topological hyperbolic lattices also correspond to regular magnetic graphs with novel {$p,q$} configurations, enabled by an effectively implemented negative curvature in the plane[11]. By contrast, topological quasicrystals[4-6] and hyperuniform or amorphous topological materials[7-10] correspond to irregular magnetic graphs, yet retain the aperiodic or statistical order required to support gapped spectra.

This network-based perspective naturally raises a question: can topological wave phenomena be realized in arbitrary, or more complex, networks? Despite substantial generalizations of topological platforms, the variants considered so far—spanning periodic[3], quasiperiodic[4-6], non-Euclidean[11,15,16], higher-order[17,18] or matrix-valued[19,20] hopping, and disordered media[7,8]—still represent only a narrow subset of network topologies, even though they are built on markedly different conceptual frameworks. The broader landscape of network topologies[21,22], ranging from regular to random networks and their intermediate regimes, motivates the next step in topological physics—the realization of topological phases in highly complex networks. Especially, heavy-tailed networks, exhibiting degree-dependent sensitivity of signal transport and small-world characteristics[23-27], may enable new ways to tailor nontrivial topological phenomena, which are intrinsically robust owing to topological protection.

Here, we extend topological phenomena to heavy-tailed networks. Developing a criterion for implementing 2D topological physics in complex networks, we investigate the Apollonian network[28] as a representative example, which possesses a fractal-like spectrum with abundant flat bands. To probe topological phases of the network, we develop a deterministic gauge-assignment protocol for the tailored flux distribution, thereby achieving the flux-dependent spectrum of the topological Apollonian network, which we term the "Apollonian butterfly." By measuring its topological properties using spectral localizers[29], we demonstrate that the topological behaviour of this heavy-tailed network is primarily driven by low-degree, non-hub nodes, offering a striking parallel to the controllability of complex networks[30]. This result establishes a framework for extending topological phenomena to high-degree networks, paving the way for controlling topological waves via network connectivity.

## Results

**Planar networks for 2D topology**

To generalize 2D topological phenomena to complex networks, we consider an extension of the Hofstadter model[3] by realizing a uniform magnetic flux across 2D plaquettes, as established in previous implementations[4-8,11]. In these implementations, the uniform magnetic flux is defined by the Peierls phase $\theta = \oint_{\partial S} \mathbf{A} \cdot d\mathbf{l}$ across a plaquette $S$, where $\mathbf{A}$ is the gauge field along the boundary links $\partial S$. As a natural extension of the Hofstadter model, such generalized implementations require an unambiguously defined magnetic flux $\theta$ for each plaquette, which is straightforward in simple graphs composed of nonoverlapping, low-degree coupling links (Fig. 1a), as considered in previous studies[4-8,11].

However, for graphs with more intricate connectivity (Fig. 1b), one must verify whether the network supports a unique flux for each plaquette. If the nodes can be embedded in a 2D plane without link crossings (Fig. 1c)—defining a planar graph network— the system can be designed to support unique flux per plaquette, enabling a natural extension of 2D topological physics to complex networks. Conversely, when a network is nonplanar—such as the complete graph on five vertices according to Kuratowski's theorem[31] (Fig. 1d)—the realization of 2D topological physics is no longer feasible, because the geometric definition of a plaquette becomes ill-defined.

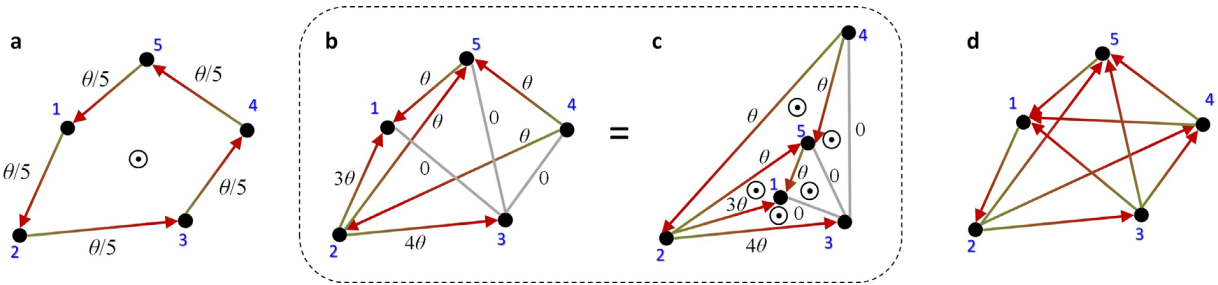

**Fig. 1. Planar networks for 2D topological physics**. **a,** A simple network without link crossings. **b,** A more intricate network with initial link crossings from higher-order couplings, and **c,** its transformed planar embedding where crossings are removed. **d,** A nonplanar complete graph, which precludes a uniquely defined magnetic flux across its plaquettes.

**Apollonian wave networks**

Given that planarity is a prerequisite for realizing 2D topological physics, investigating 2D topological phases in heavy-tailed environments naturally entails the utilization of heavy-tailed planar networks. As a representative example of such architectures[28,32-34], we employ the Apollonian networks[28,35-37] as our platform. This network is generated via a deterministic evolution rule—recursive face subdivision (Supplementary Note S1)—which yields a maximal planar graph featuring a hierarchical structure of nodes. Figure 2a illustrates the Apollonian network obtained after five epochs ($g = 5$), exhibiting differentiated node degrees for unweighted links.

Owing to its deterministic nature, the heavy-tailed properties of the Apollonian network can be examined analytically. At the $e$-th epoch of evolution, $3^{e-1}$ nodes are newly generated, each of which acquires a node degree of $3 \times 2^{g-e}$ after completing the $g$-epoch generation ($g \geq e$). Because the first three nodes possess a degree of $2^g + 1$, the probability mass function (PMF) at the $g$-th epoch is given by

$$p_g(x_e) = \begin{cases} \dfrac{6}{3^g + 5} & (e = 0) \\ \dfrac{2 \times 3^{e-1}}{3^g + 5} & (e = 1, 2, ..., g) \end{cases}, \quad (1)$$

where $x_0 = 2^g + 1$ and $x_e = 3 \times 2^{g-e}$ for $e = 1, 2, …, g$. Consequently, the Apollonian network exhibits the power-law degree distribution for sufficiently large $g$, as $p_g(x) \propto x^{-\ln 3/\ln 2}$ (Fig. 2b), thereby satisfying the condition of an extremely heavy-tailed network with hub nodes[23] (Fig. 2a). The only deviation from the power law arises from the three initial nodes (blue dashed circles in Fig. 2a,b), which correspond to the boundary elements of the system.

In terms of generalizing topological phases in Euclidean[1,2] and non-Euclidean[11] planar geometries to network science, we focus on the unweighted Apollonian network. When assigning a node as a localized state and a link as the interstate coupling, the target configuration can be implemented via distance-independent coupling, which has been experimentally realized using waveguide loop couplers in integrated photonics[15,38] (Fig. 2c), capacitive junctions in circuit quantum electrodynamics[39] (Fig. 2d) or high-frequency circuits[16,40], or long-range coupling via epsilon-near-zero media[41]. Based on this unweighted link model, the network can be regarded as a regular graph composed of equilateral triangle faces, which possesses the following effective Gaussian curvature $K_n$ at the $n$-th node:

$$K_n = \frac{\pi}{S}\left(\frac{6}{q_n} - 1\right), \tag{2}$$

where $q_n$ denotes the node degree and $S$ is the area of a triangle face. Consequently, the network represents a non-Euclidean plane with spatially inhomogeneous curvature—hyperbolic regions near hub nodes, approximately Euclidean regions around intermediate nodes, and elliptic regions around peripheral nodes (Fig. 2e).

Notably, the structures for distance-independent coupling can be modified to independently engineer gauge fields along each network link to achieve a tailored flux distribution[11,15,16,38]. When considering an identical single mode at each node—for example, a pseudospin mode of an identical unit resonator[38]—wave dynamics on the Apollonian network are governed by the following tight-binding Hamiltonian:

$$H = t\sum_{\langle m,n \rangle}\left(e^{-i\varphi_{mn}} a_m^\dagger a_n + h.c.\right), \tag{3}$$

where $a_m^\dagger$ (or $a_m$) is the creation (or annihilation) operator at the $m$-th node, $t$ is the coupling strength between the sites, $\varphi_{mn}$ is the additionally acquired phase from the $n^{\text{th}}$ to $m^{\text{th}}$ sites via the link gauge, the pair ⟨m,n⟩ denotes the connected nodes, and *h.c.* denotes the Hermitian conjugate. Throughout the following discussion, we set $t = 1$.

Figures 2f-2i present the sorted eigenvalues $E$ of the topologically trivial eigenvalue problem $H|\psi\rangle = E|\psi\rangle$ with $\varphi_{mn} = 0$ at $g = 9$ (Supplementary Note S2 for evolving $g$). When plotted against the logarithmically scaled normalized mode number, the gapped spectrum alternates between extended plateaus (grey) that correspond to flat-band manifolds and intervening predominantly dispersive regions (yellow) that are interspersed with fine, step-like micro-plateaus. The successive magnifications in Figs. 2g-2i reveal a hierarchical, scale-invariant organization of plateaus and dispersive bands: similar step-like structures recur over fractional intervals of the mode index. Overall, the flat bands occupy approximately one third of the eigenmodes, while the remaining modes populate dispersive bands, yielding a distinctive spectrum of the topologically trivial Apollonian network characterized by self-similarity and multiple gaps.

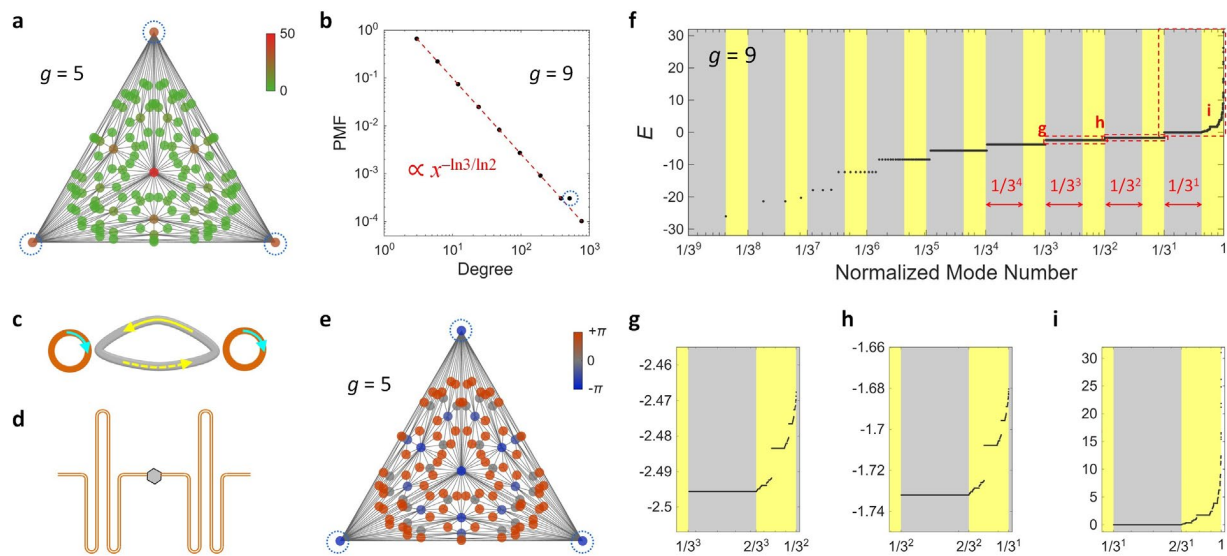

**Fig. 2. Apollonian wave networks and their topologically trivial spectrum**. **a,** Deterministic Apollonian network after five epochs, showing differentiated node degrees (colour bar) for unweighted links. The three boundary nodes are highlighted with dashed circles. **b,** Degree PMF at $g = 9$ obeying $p_g(x) \propto x^{-\ln 3/\ln 2}$, with deviations only at the initial nodes (dashed circle). **c,d,** The possible configurations for distance-independent coupling implemented via waveguide-loop couplers (**c**) and capacitive junctions (**d**). **e,** Effective Gaussian curvature $K_n$ (colour bar) mapped on the $g = 5$ network, indicating hyperbolic hubs and elliptic peripheries. **f,** Sorted eigenvalues $E$ as a function of the normalized mode number, exhibiting macroscopic flat-band plateaus (grey) and predominantly dispersive intervals (yellow). **g-i,** Enlarged figures for g-i regions (red dashed boxes) in **f**.

**Design of magnetic fluxes**

In the conventional Hofstadter model[3], a magnetic flux per plaquette is implemented through Peierls phases associated with a convenient gauge choice, such as the Landau gauge. The key simplification in the Hofstadter model on a periodic lattice is that, for rational flux, one can introduce a magnetic unit cell and analyse the problem within a Bloch-theorem framework, thereby enabling a transparent design of the gauge-phase pattern especially for a uniform flux distribution. By contrast, in the absence of translational periodicity, there is no canonical gauge pattern tied to lattice vectors. Therefore, designing spatially tailored flux profiles becomes an inverse problem on a general planar network, constrained by its topology. Although the recipe for hyperbolic lattices under uniform magnetic fluxes have been proposed[11], the method based on the regular tiling of the lattice is not directly transferable to our Apollonian networks with strongly inhomogeneous node degrees.

To address this challenge, we develop a deterministic gauge-assignment algorithm that guarantees the desired Peierls phase around every triangular face of the Apollonian network. This algorithm is structured by exploiting network planarity—an interior link is shared by two faces and thus couples their flux constraints, whereas an outermost boundary link participates in only a single boundary face. Therefore, boundary links retain the largest residual gauge freedom, enabling final-step corrections without altering flux constraints already satisfied in the interior.

From this perspective, we implement an evolutionary gauge-assignment scheme based on the dual graph—a graph that has a vertex for each face of a given graph—of the Apollonian network. After placing nodes at face centres and connecting adjacent faces via a dual-graph link (Fig. 3a), we employ a multi-source breadth-first search (BFS)[42] starting from the three boundary faces to obtain a distance ordering (Fig. 3b). Following this ordering, we sequentially assign link phases from faces farthest from the boundary toward the boundary (Fig. 3c), ensuring that each newly processed face does not alter the flux constraints already satisfied by previously assigned faces. This procedure enables flexible synthesis of tailored flux landscapes, including uniform flux distributions (Figs. 3d and 3e) as well as random flux profiles (Fig. 3f).

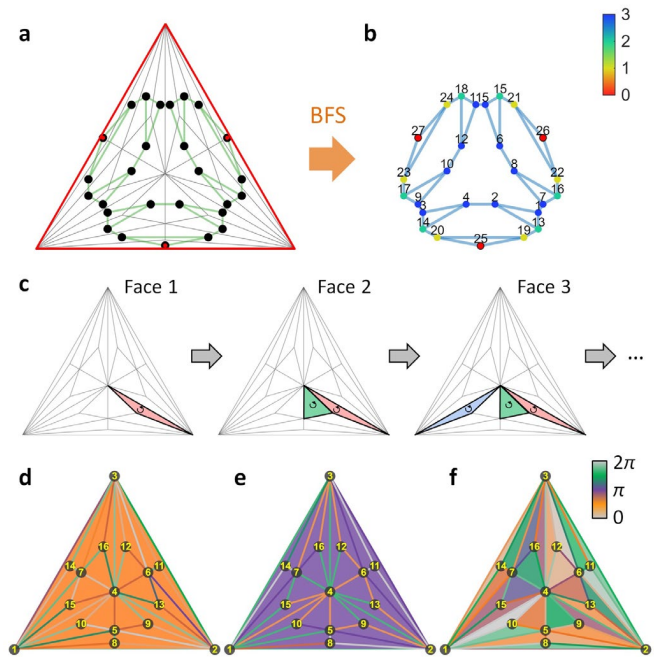

**Fig. 3. Evolutionary gauge design in Apollonian networks. a,** Dual graph construction: face centres define nodes (black circles) and adjacent faces compose a link (green lines). The boundary face nodes and boundary links are highlighted by red dots and lines, respectively. **b,** Multi-source BFS layering from the three boundary faces. Smaller indices denote faces farther from the boundary. Colours represent the minimal distance from the boundaries. **c,** Evolutionary protocol that assigns link phases face by face from the deepest interior. Distinct face colours represent the tailored flux values. **d-f**, Resulting fluxes (face colours) and gauge-field (line colours) distributions for uniform ($\theta = \pi/4$ in **d** and $\theta = \pi$ in **e**) and random (**f**) fluxes. We set $g = 3$ for clarity.

**Topological Apollonian networks**

Based on the design in Fig. 3, we investigate topological Apollonian networks under a uniform flux $\theta$. As a counterpart of the Hofstadter butterfly[3], we compute the "Apollonian butterfly"—the eigenenergy spectrum $E$ as a function of $\theta$ (Fig. 4a; Supplementary Note S3 for evolving $g$)—which obeys the spectral symmetries, $\sigma(E(\theta)) = \sigma(E(-\theta))$ and $\sigma(E(\theta)) = -\sigma(E(\theta + \pi))$, where $\sigma(E(\theta)) = \{E(\theta)\}$ denotes the set of eigenvalues at flux $\theta$. Notably, these symmetries reflect both topological and microstructural features of the Apollonian network. First, $\sigma(E(\theta)) = \sigma(E(-\theta))$—a property shared with the Hofstadter butterfly—follows from the time-reversal transformation, $TH(\theta)T^{-1} = H(-\theta)$, where $T$ is the time-reversal operator. This condition implies that time-reversal symmetry is generally broken, because $H(\theta) \neq H(-\theta)$ for generic $\theta$.

We note that the distinctive spectral feature of the Apollonian butterfly is $\sigma(E(\theta)) = -\sigma(E(\theta + \pi))$. This property originates from the microstructural characteristics of the Apollonian network—a maximal planar graph obtained by triangulation—which leads to the flux-shifted chiral symmetry, $\Gamma H(\theta) \Gamma^{\dagger} = -H(\theta + \pi)$, where $\Gamma$ is a chiral-like diagonal unitary operator (Supplementary Note S4). While $\Gamma$ depends on the gauge-assignment algorithm described in Section IV, the underlying microstructural origin is in sharp contrast to the chiral symmetry of the Hofstadter butterfly, which arises from its bipartite network.

To further explore the topological properties of the Apollonian network, we evaluate a real-space topological invariant. Because the Bloch theorem is not applicable, we characterize the topological phase in real space using the spectral localizer $L(m,E;\kappa)$[29], defined as

$$L(m, E; \kappa) = (H - EI) \otimes \sigma_z + \kappa \left[ (X - x_m I) \otimes \sigma_x + (Y - y_m I) \otimes \sigma_y \right], \tag{4}$$

where $I$ is the identity matrix, $(x_m, y_m)$ is the 2D real-space coordinate of the $m$-th node, $X$ and $Y$ are the position operators, $\sigma_{x,y,z}$ are the Pauli matrices, and $\kappa$ is the Hamiltonian-position scaling parameter. From the eigenvalues $\{\lambda\}$ of $L(m,E;\kappa)$, we compute two local measures: the local Chern marker $C(m,E;\kappa)$, defined as

$$C(m, E; \kappa) = \frac{1}{2}\mathrm{sig}(L(m, E; \kappa)), \tag{5}$$

and the local gap $\mu(m,E;\kappa)$ defined as

$$\mu(m, E; \kappa) = \min\left[\{|\lambda|\}\right], \tag{6}$$

where $\mathrm{sig}(L(m,E;\kappa))$ denotes the signature of $L$—the difference between the numbers of positive and negative eigenvalues. In our calculation, we determine $\kappa$ following the recipe in the previous work[29] (Supplementary Note S5).

As described in Supplementary Note S3, the spectral geometry of the Apollonian butterfly converges to a characteristic shape first near $E = 0$ as $g$ increases. This observation suggests that the inhomogeneous bulk properties of the finite-$g$ Apollonian butterfly are reflected in the near-zero-energy regime. Therefore, in analysing the topological measures, we focus on the region $|E|^2 \leq 3$—the spectral window between the two flat bands near $E = 0$—to probe the inhomogeneous bulk characteristics of the topological Apollonian network. Figure 4b shows the nontrivial Chern marker $C(m,E;\kappa)$ across the network, where only $\theta$-$E$ points with nontrivial Chern markers are coloured. We note that the regions $0 < \theta < \pi$ and $\pi < \theta < 2\pi$ exhibit opposite signs of $C$, consistent with the time-reversal mapping $\theta \to -\theta$ and the associated relation $C(\theta) = -C(-\theta)$, as in the conventional Hofstadter butterfly. In contrast, $C = 0$ near $\theta = 0$ and $\theta = \pi$, indicating that time-reversal symmetry is effectively restored at these flux values.

Figure 4c presents another critical topological measure—the average local gap $\langle \mu \rangle$ computed over the nontrivial nodes—quantifying how well the topological phase is spectrally protected by a locally gapped spectrum. Consistent with the Chern-marker map in Fig. 4b, $\langle u \rangle$ forms extended gapped domains within the spectral window near $E = 0$, while it diminishes near $\theta = 0$ and $\theta = \pi$ where topological character changes. Notably, the pattern of $\langle u \rangle$ follows the same spectral symmetries of $\sigma(E(\theta)) = \sigma(E(-\theta))$ and $\sigma(E(\theta)) = -\sigma(E(\theta + \pi))$, indicating that the stability of the topological response is controlled jointly by the time reversal transformation and the triangulation-induced microstructure.

Along with the spectral results in Figs. 4b and 4c, we examine the real-space distribution of the local Chern marker $C$ to assess the impact of the highly inhomogeneous node degrees in the Apollonian network. Figures 4d-g show $C(m,E;\kappa)$ at selected values of $\theta$ and $E$ indicated in Fig. 4c. In the region where $\langle u \rangle$ is largest (Fig. 4d)—that is, the most robust topological regime—the bulk region away from the network boundary exhibits a nontrivial topological phase, as in other finite topological systems[43,44]. Notably, the heavy-tailed nature of the Apollonian network becomes apparent as $\langle u \rangle$ decreases (Figs. 4e to 4g). Comparing Figs. 4e-g with Fig. 2a demonstrate that high-degree nodes—hubs—lose their nontrivial topological character more readily. This behaviour is consistent with a hallmark of heavy-tailed networks: fragile hubs and robust peripheries[22,25].

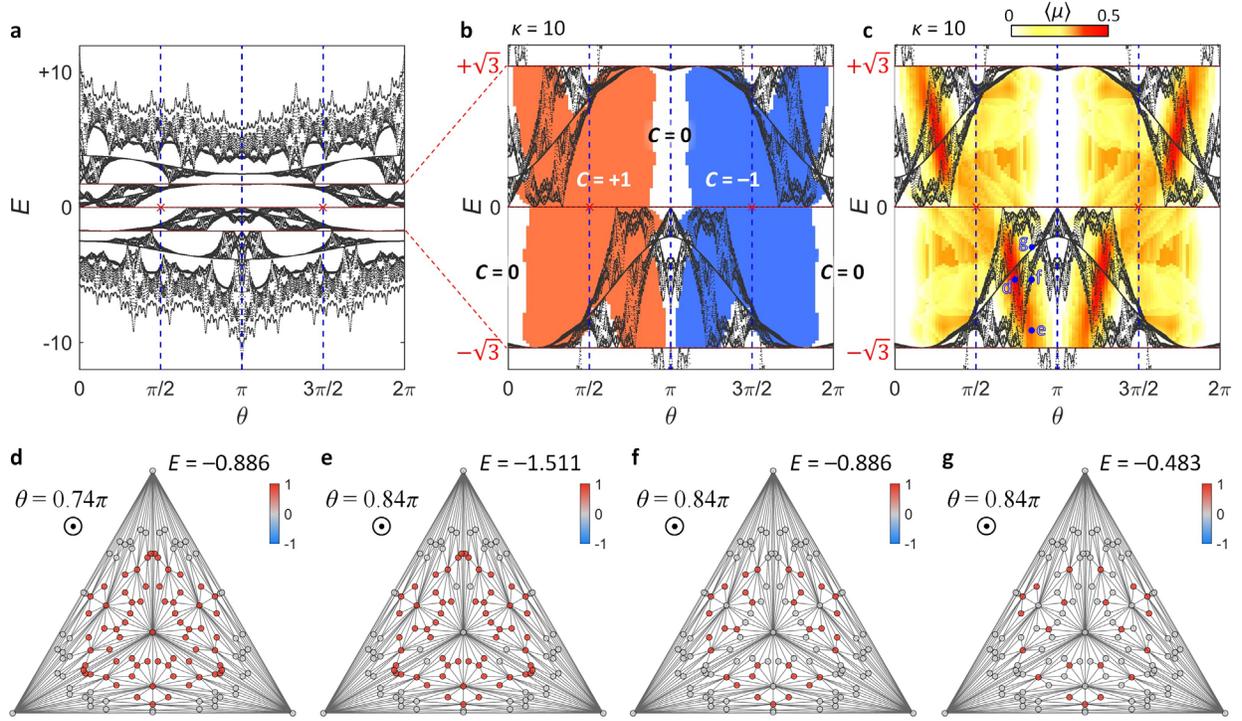

**Fig. 4. Apollonian butterfly and topological measures. a,** Apollonian butterfly obtained by plotting the eigenstates (black dots) in the $\theta$-$E$ plane. **b,** Local Chern marker $C$: coloured pixels indicate a nontrivial $C$ in at least one node, whereas white pixels indicate a trivial $C$ over the entire network. **c,** Averaged local gap $\langle\mu\rangle$, calculated over the nontrivial nodes. Blue dashed lines and red crosses in **a-c** are shown to highlight the symmetry conditions, $\sigma(E(\theta)) = \sigma(E(-\theta))$ and $\sigma(E(\theta)) = -\sigma(E(\theta + \pi))$. **d-g,** Distributions of $C(m,E;\kappa)$ at selected values of $\theta$ and $E$, depicted in **c**. $\kappa = 10$ in **b-g**. We set $g = 5$ in all cases.

**Topological controllability**

In studies of heavy-tailed networks, the impact of hubs is multifaceted[25,30]. As demonstrated also in our wave network example (Fig. 4), various intriguing features of such networks originate from their differentiated robustness against perturbations to hubs and peripheral nodes; signal transport is particularly fragile when hub nodes are perturbed[22,25]. However, from the perspective of system controllability, this well-known picture becomes reversed. Highly connected hubs tend to impede efficient controllability, whereas driver nodes are more frequently found among low-degree nodes[30,45]—as if a small gear with fewer connections, rather than a large one with many, were orchestrating the entire system. Motivated by these intriguing results, we examine the impact of specific nodes on the spectral control of topological Apollonian networks.

We consider a one-dimensional (1D) lattice of Apollonian networks (Fig. 5a), which is modelled using the Bloch theorem with crystal momentum $\beta$. Each network is subjected to a uniform magnetic flux $\theta = 0.74\pi$, for which a pronounced local gap appears near $E = -0.886$ (Fig. 4d). Under this topological setting, the dispersion bands of the 1D lattice become nonreciprocal, as shown in Figs. 5b-5e (yellow lines).

To examine the controllability of the system bands, we classify nodes into four groups according to their node degrees: 3, 6, 12, and {24, 33, 48}, comprising 81, 27, 9, and 7 nodes, respectively. For a fair comparison, we randomly select five nodes from each group and apply onsite perturbations $\delta a_m^\dagger a_m$ to the Hamiltonian, where each $\delta$ is independently sampled from a zero-mean normal distribution with standard deviation $\delta_0 = 0.30$ (Supplementary Note S6 for other $\delta_0$). The black dots in Figs. 5b-5e show the corresponding perturbed bands for an ensemble of random realizations in each group.

While the relative robustness near $E = -0.886$ confirms the validity of the local gap described in Fig. 4c, Figs. 5b-5e further demonstrate that the driver nodes in topological Apollonian networks are generally peripheral in the near-zero-energy regime. Notably, the node degree determines which part of the spectrum is most effectively driven as supported by the distinct eigenstate profiles in Figs. 5f-5h; degree-3 nodes primarily affect the high $E$ region (Fig. 5b), degree-6 nodes mainly affect the low $E$ region (Fig. 5c), and degree-12 nodes selectively influence the high local-gap region (Fig. 5d). Remarkably, hub nodes with degrees 24, 33, and 48 exhibit little ability to control the bands in the spectral region of interest, which is partly counterintuitive given the well-known fragility of hubs. This behaviour can be understood as a diminished marginal impact of hubs arising from their extreme connectivity. Such dense connectivity promotes destructive interference, as evidenced by the unexcited hub nodes in Figs. 5f-5h, which is also consistent with the difficulty of sustaining topological characteristics shown in Figs. 4e-4g.

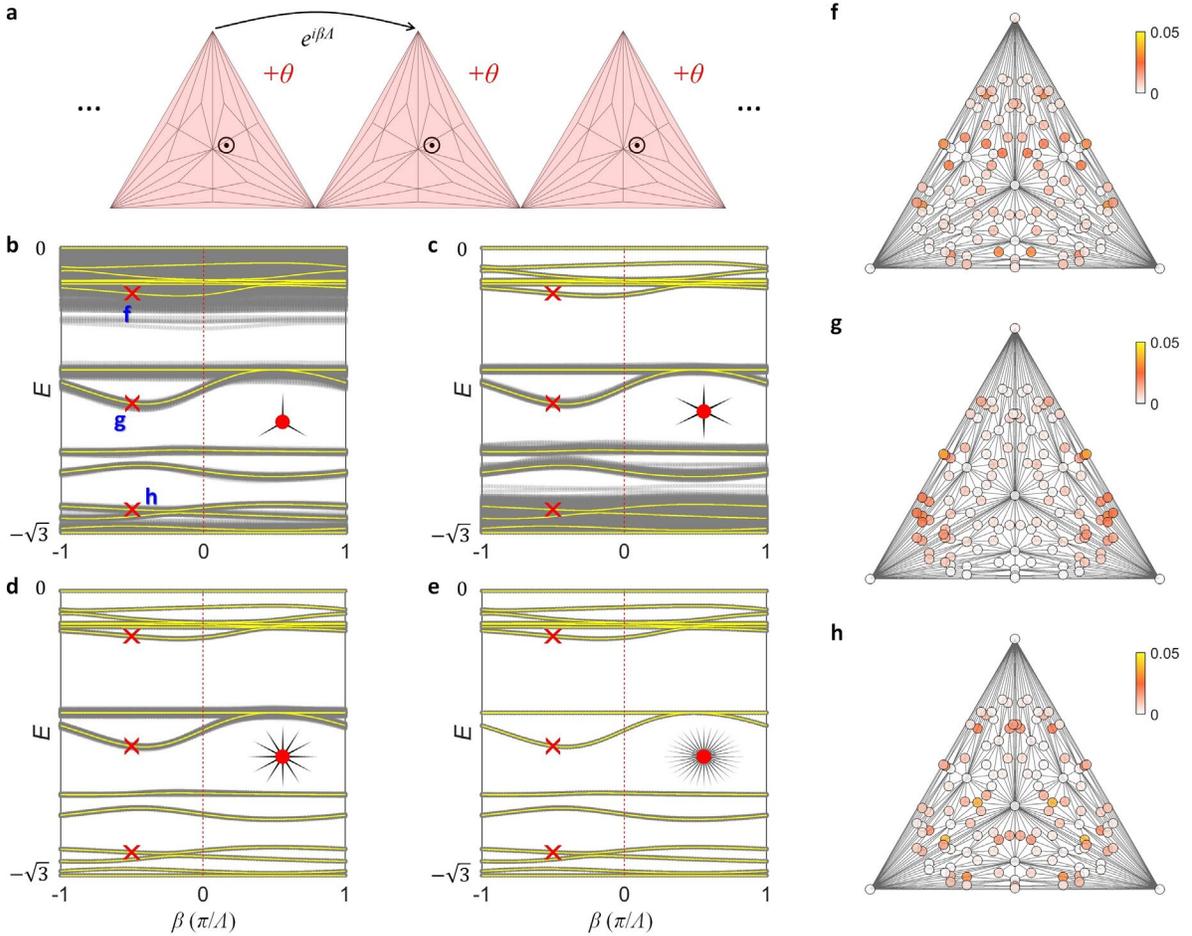

**Fig. 5. Topological band controllability**. **a,** Schematic of a 1D periodic array of Apollonian unit networks under a uniform magnetic flux $\theta$. $\Lambda$ denotes the periodicity. **b-e,** Nonreciprocal Bloch bands $E(\beta)$ at $\theta = 0.74\pi$ (yellow lines) and their sensitivity to node perturbations applied to different degree classes: the perturbations of node degree 3 (**b**), 6 (**c**), 12 (**d**), and hub nodes with degrees {24, 33, 48} (**e**). The overlaid black dots show the perturbed eigenvalues for an ensemble of 100 random realizations obtained by perturbing five randomly chosen nodes per group with normal noise of $\delta_0 = 0.30$. The red marker illustrates a perturbed node with links. **f-h,** The intensity of the wavefunctions $|\psi(m)|^2$ for the unperturbed states depicted in **b** (red crosses): $E = -0.275$ (**f**), $-0.944$ (**g**), and $-1.592$ (**h**) at $\beta = -0.5\pi/\Lambda$. We set $g = 5$ in all cases.

## Discussion

Although we have established a concrete route to implementing 2D topological physics in the Apollonian network, several questions remain resolved. First, owing to the extremely short graph distances and dense hierarchical connectivity, it is nontrivial to isolate conventional edge transport within a single finite network. Notably, this ambiguity is related to hub localization at the spectral boundary (Supplementary Note S7), which suggests that hub nodes serve as the spatial boundary characterizing the finiteness of the geometry. Therefore, a natural next step is to promote the Apollonian network to the unit cell of a larger 2D architecture, in which an array of hub elements forms the interface for evaluating the bulk-boundary correspondence. Second, our topological characterization has focused primarily on the spectral window near $E = 0$—the regime featuring the inhomogeneous bulk properties. The Apollonian butterfly contains rich high- and low-energy structures with additional gaps and flat bands, where one can systematically search for higher-magnitude topological invariants for high capacity information transport[18].

From a network-science viewpoint, the topological Apollonian network should be regarded as only one representative of planar magnetic graphs. We envisage the use of random Apollonian networks[46] or planarity-constrained growth models[47] to disentangle which topological features are tied to highly symmetric triangulation, which to heavy tails, and which to algorithmic gauge assignment. Beyond 2D geometry, the recent surge of interest in higher-dimensional topology[48] motivates an analogous approach for topological networks embeddable in higher-dimensional manifolds.

In conclusion, we have introduced a framework for realizing topological physics in a heavy-tailed planar network. By devising an evolutionary gauge-assignment scheme, we constructed Apollonian networks under tailored flux distributions and computed the Apollonian butterfly under a uniform flux. We reveal spectral symmetries induced by time-reversal transformation and microstructural triangulation. Real-space measures of local Chern markers and local gaps connect topological robustness to the inhomogeneous connectivity of the network. While the Apollonian network exhibits topological behaviour similar to that of regular lattices in the large local-gap regime, its heavy-tailed nature becomes evident as the local gap decreases, revealing the fragility of hub nodes in maintaining nontrivial topological characteristics. At the same time, this observation links topological protection to network controllability, elucidating why low-degree nodes become driver nodes. Our results, obtained from the first-ever implementation of 2D topological physics in high-degree ($> 40$), inhomogeneous networks, extend topological physics into the realm of complex networks.

## Data availability

The data used in this study are available from the corresponding authors upon request and can also be accessed by running the code provided as Supplementary Code S1.

## Code availability

The code used in this study is developed by the authors, and is provided as Supplementary Code S1 to reproduce all data presented in this paper.

## Acknowledgements

We acknowledge financial support from the National Research Foundation of Korea (NRF) through the Basic Research Laboratory (No. RS-2024-00397664), Innovation Research Center (No. RS-2024-00413957), Pilot and Feasibility Grants (No. RS-2025-19912971), Young Researcher Program (No. RS-2025-00552989), and Midcareer Researcher Program (No. RS-2023-00274348), all funded by the Korean government. This work was supported by Creative-Pioneering Researchers Program through Seoul National University. We also acknowledge an administrative support from SOFT foundry institute.

## Author Contributions

All authors contributed equally to conceiving the idea, developing the code, discussing the results, and preparing the final manuscript.

## Competing Interests

The authors have no conflicts of interest to declare.

## Additional information

**Correspondence and requests for materials** should be addressed to S.Y., X.P., or N.P.

# Figure Legends

**Fig. 1. Planar networks for 2D topological physics**. **a,** A simple network without link crossings. **b,** A more intricate network with initial link crossings from higher-order couplings, and **c,** its transformed planar embedding where crossings are removed. **d,** A nonplanar complete graph, which precludes a uniquely defined magnetic flux across its plaquettes.

**Fig. 2. Apollonian wave networks and their topologically trivial spectrum**. **a,** Deterministic Apollonian network after five epochs, showing differentiated node degrees (colour bar) for unweighted links. The three boundary nodes are highlighted with dashed circles. **b,** Degree PMF at $g = 9$ obeying $p_g(x) \propto x^{-\ln3/\ln2}$, with deviations only at the initial nodes (dashed circle). **c,d,** The possible configurations for distance-independent coupling implemented via waveguide-loop couplers (**c**) and capacitive junctions (**d**). **e,** Effective Gaussian curvature $K_n$ (colour bar) mapped on the $g = 5$ network, indicating hyperbolic hubs and elliptic peripheries. **f,** Sorted eigenvalues $E$ as a function of the normalized mode number, exhibiting macroscopic flat-band plateaus (grey) and predominantly dispersive intervals (yellow). **g-i**, Enlarged figures for g-i regions (red dashed boxes) in **f**.

**Fig. 3. Evolutionary gauge design in Apollonian networks. a,** Dual graph construction: face centres define nodes (black circles) and adjacent faces compose a link (green lines). The boundary face nodes and boundary links are highlighted by red dots and lines, respectively. **b,** Multi-source BFS layering from the three boundary faces. Smaller indices denote faces farther from the boundary. Colours represent the minimal distance from the boundaries. **c,** Evolutionary protocol that assigns link phases face by face from the deepest interior. Distinct face colours represent the tailored flux values. **d-f**, Resulting fluxes (face colours) and gauge-field (line colours) distributions for uniform ($\theta = \pi/4$ in **d** and $\theta = \pi$ in **e**) and random (**f**) fluxes. We set $g = 3$ for clarity.

**Fig. 4. Apollonian butterfly and topological measures. a,** Apollonian butterfly obtained by plotting the eigenstates (black dots) in the $\theta$-$E$ plane. **b,** Local Chern marker $C$: coloured pixels indicate a nontrivial $C$ in at least one node, whereas white pixels indicate a trivial $C$ over the entire network. **c,** Averaged local gap $\langle \mu \rangle$, calculated over the nontrivial nodes. Blue dashed lines and red crosses in **a-c** are shown to highlight the symmetry conditions, $\sigma(E(\theta)) = \sigma(E(-\theta))$ and $\sigma(E(\theta)) = -$

$\sigma(E(\theta + \pi))$. **d-g,** Distributions of $C(m,E;\kappa)$ at selected values of $\theta$ and $E$, depicted in **c**. $\kappa = 10$ in **b-g**. We set $g = 5$ in all cases.

**Fig. 5. Topological band controllability**. **a,** Schematic of a 1D periodic array of Apollonian unit networks under a uniform magnetic flux $\theta$. $\Lambda$ denotes the periodicity. **b-e,** Nonreciprocal Bloch bands $E(\beta)$ at $\theta = 0.74\pi$ (yellow lines) and their sensitivity to node perturbations applied to different degree classes: the perturbations of node degree 3 (**b**), 6 (**c**), 12 (**d**), and hub nodes with degrees {24, 33, 48} (**e**). The overlaid black dots show the perturbed eigenvalues for an ensemble of 100 random realizations obtained by perturbing five randomly chosen nodes per group with normal noise of $\delta_0 = 0.30$. The red marker illustrates a perturbed node with links. **f-h,** The intensity of the wavefunctions $|\psi(m)|^2$ for the unperturbed states depicted in **b** (red crosses): $E = -0.275$ (**f**), $-0.944$ (**g**), and $-1.592$ (**h**) at $\beta = -0.5\pi/\Lambda$. We set $g = 5$ in all cases.

# Supplementary Information for "Topological Heavy-Tailed Networks"

Sunkyu Yu[1†], Xianji Piao[2§], and Namkyoo Park[3*]

[1]Intelligent Wave Systems Laboratory, Department of Electrical and Computer Engineering, Seoul National University, Seoul 08826, Korea

[2]Wave Engineering Laboratory, School of Electrical and Computer Engineering, University of Seoul, Seoul 02504, Korea

[3]Photonic Systems Laboratory, Department of Electrical and Computer Engineering, Seoul National University, Seoul 08826, Korea

E-mail address for correspondence: [†]sunkyu.yu@snu.ac.kr, [§]piao@uos.ac.kr, [*]nkpark@snu.ac.kr

**Note S1. Generation of Apollonian networks**

**Note S2. $g$-dependent spectra in topologically trivial Apollonian networks**

**Note S3. $g$-dependent Apollonian butterflies**

**Note S4. Flux-shifted chiral symmetry**

**Note S5. Hamiltonian-position scaling parameter**

**Note S6. Dependence on perturbation strength**

**Note S7. Hub localization**

**Note S1. Generation of Apollonian networks**

Figure S1 illustrates the construction of the $g$-th epoch Apollonian network. Starting from a seed triangle ($g = 0$), at each step every triangular face in the $g$-th epoch is subdivided in the $(g+1)$-th epoch by adding a new node—totally, $3^{g-1}$ new nodes in the $g$-th epoch network—at its centroid and connecting it to the three vertices of the face.

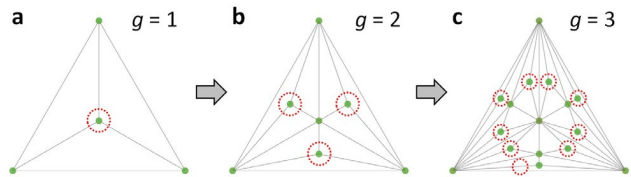

**Fig. S1. Evolutionary design of Apollonian networks.** The networks in the $g = 1$ (**a**), $g = 2$ (**b**), and $g = 3$ (**c**) epochs.

**Note S2. *g*-dependent spectra in topologically trivial Apollonian networks**

Figure S2 indicates that the energy spectrum develops a hierarchical structure as the generation *g* increases. While only a limited number of discrete eigenvalues and short plateaus are visible for lower *g*, increasing *g* leads to more densely populated and organized spectra, yielding broad flat-band plateaus separated by dispersive intervals. This tendency demonstrates that the recursive growth of the Apollonian network enhances spectral division while preserving the coexistence of flat bands (grey) and dispersive bands (yellow). Therefore, Fig. S2 suggests that the generation number not only controls the total spectral complexity but also derives the analytical balance between flat-band localization and dispersive transport.

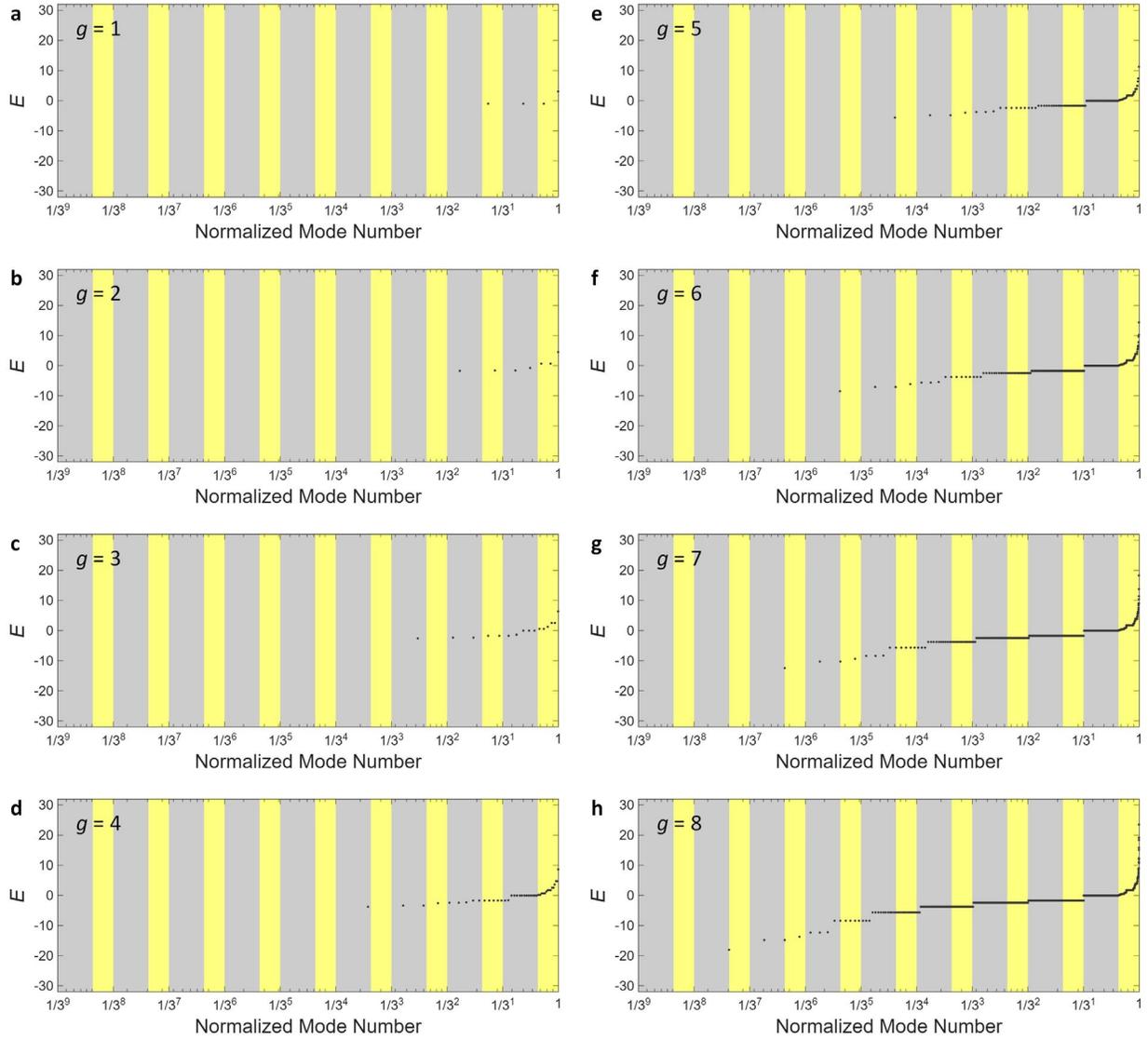

**Fig. S2. Generation dependency of the energy spectra.** Sorted eigenvalues $E$ as a function of the normalized mode number for different $g$ values. Macroscopic flat-band plateaus (grey) and predominantly dispersive intervals (yellow) are indicated exactly as in Fig. 2f of the main text.

**Note S3. *g*-dependent Apollonian butterflies**

Figure S3 demonstrates that the structure of the Apollonian butterfly evolves rapidly with the generation number *g*, transforming from a relatively simple set of dispersive bands at low *g* into a highly intricate, multilayered pattern for higher *g*. As *g* increases, the number of bands grows rapidly, indicating the emergence of hierarchical level splitting associated with the microstructure of Apollonian geometry. At the same time, the enlarged views of the energy window $|E|^2 \leq 3$ reveal that the central butterfly structure is already visible at intermediate generations ($g \cong 3$) and becomes increasingly sharp for $g \geq 5$. Notably, prominent gaps and symmetry-related features remain robust throughout the generation growth, implying that the given butterfly is not a finite-size artifact but an intrinsic consequence of the topological characteristics of the Apollonian network.

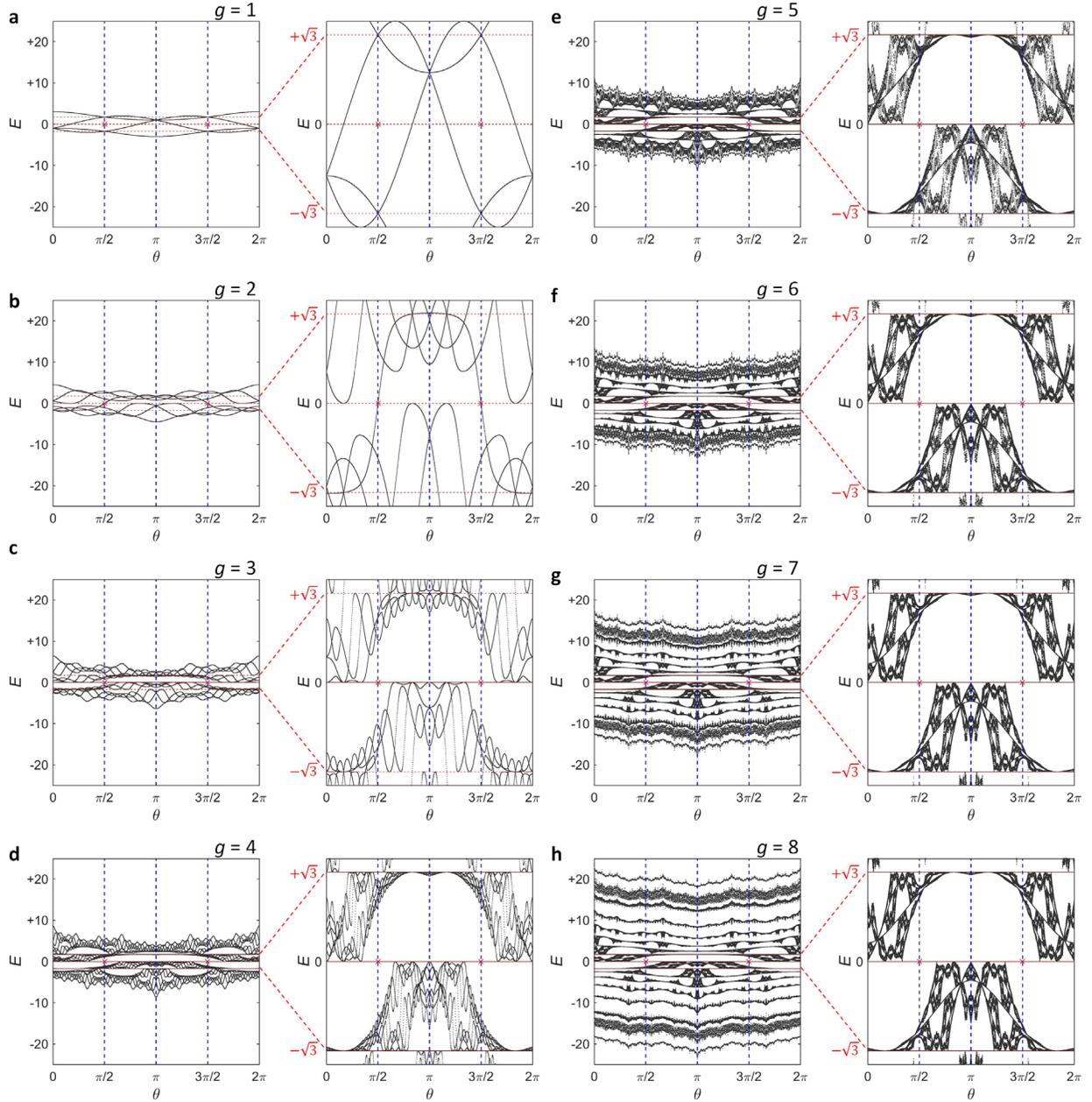

**Fig. S3. Generation dependence of the Apollonian butterfly.** For each subfigure, the left and right panels show the full $\theta$-$E$ plane and an enlarged view of the region $|E|^2 \leq 3$, respectively.

**Note S4. Flux-shifted chiral symmetry**

Assuming that a uniform flux is implemented across the entire Apollonian network, Eq. (3) in the main text can be rewritten as

$$H(\theta) = t \sum_{\langle m,n \rangle} \left( e^{-i\varphi_{mn}(\theta)} a_m^\dagger a_n + h.c. \right). \tag{S1}$$

where $\varphi_{mn}(\theta)$ is designed to realize the plaquette flux $\theta$. From Eq. (S1), we obtain

$$\begin{aligned}-H(\theta) &= -t \sum_{\langle m,n \rangle} \left( e^{-i\varphi_{mn}(\theta)} a_m^\dagger a_n + h.c. \right) \\ &= t \sum_{\langle m,n \rangle} \left( e^{-i[\varphi_{mn}(\theta)+\pi]} a_m^\dagger a_n + h.c. \right).\end{aligned} \tag{S2}$$

Because of the microstructural characteristics of the Apollonian network—a maximal planar graph obtained by triangulation—adding $\pi$ to all link phases shifts the flux through each triangular plaquette uniformly by $\pi$ (mod $2\pi$). Consequently, the term, $-H(\theta)$, is gauge-equivalent to $H(\theta + \pi)$, implying the existence of a unitary operator $U$ for $UH(\theta)U^\dagger = -H(\theta + \pi)$. This symmetry yields a spectral feature of the Apollonian butterfly, $\sigma(E(\theta)) = -\sigma(E(\theta + \pi))$.

To gain further insight into this symmetry, we restrict $U$ to a diagonal unitary operator, which satisfies $Ua_m U^\dagger = \exp(i\alpha_m)a_m$. Under this restriction, we obtain

$$UH(\theta)U^\dagger = t \sum_{\langle m,n \rangle} \left( e^{-i[\varphi_{mn}(\theta)+(\alpha_m - \alpha_n)]} a_m^\dagger a_n + h.c. \right). \tag{S3}$$

The symmetry, $UH(\theta)U^\dagger = -H(\theta + \pi)$, enforces the following relation for all links:

$$\varphi_{mn}(\theta) + (\alpha_m - \alpha_n) = \varphi_{mn}(\theta + \pi) + \pi \quad (\text{mod } 2\pi). \tag{S4}$$

When considering another diagonal unitary candidate $V$ satisfying $Va_m V^\dagger = \exp(i\beta_m)a_m$, we obtain

$$\alpha_m - \beta_m = \alpha_n - \beta_n \quad (\text{mod } 2\pi), \tag{S5}$$

for all connected $m$ and $n$. Owing to Eq. (S5), the diagonal unitary $U$ is unique up to a global phase, analogous to a chiral operator for a bipartite graph. Consequently, we can define a chiral-like diagonal operator $\Gamma$ up to a global phase, which satisfies $\Gamma H(\theta)\Gamma^\dagger = -H(\theta + \pi)$—the flux-shifted chiral symmetry of the topological Apollonian network. We note that the explicit form of $\Gamma$ is determined by the gauge-assignment algorithm according to Eq. (S4): the dependence on the relationship between $\varphi_{mn}(\theta)$ and $\varphi_{mn}(\theta + \pi)$.

**Note S5. Hamiltonian-position scaling parameter**

According to Eq. (4) in the main text, the hyperparameter $\kappa$ sets the scaling between spectral features of the Hamiltonian $H$ and the position operators $X$ and $Y$. Following the well-established procedure in the previous study [1], we determine $\kappa$ by starting from the lower bound:

$$\kappa \geq \frac{E_{gap}}{l}, \quad (S6)$$

where $l$ denotes the distance from the centre of the finite sample to its boundary, and $E_{gap}$ denotes the bulk bandgap of the Hamiltonian $H$ around the selected $E$. Although $l$ is specified explicitly as $l = \sqrt{3}/6$, the strongly inhomogeneous connectivity of the Apollonian network hinders a straightforward determination of $E_{gap}$. Because we focus on the spectral range near $E = 0$ and observe an apparent separation between neighbouring flat bands of $\Delta E = \sqrt{3}$, we empirically estimate the lower bound as $\kappa \gtrsim 5$, in excellent agreement with our numerical assessment.

Figure S4 shows the $\kappa$-dependent local gap $\mu$ over the spectral range of $|E|^2 \leq 3$, plotting $\mu(m,E;\kappa)$ for all nodes $m$ that possess a nontrivial $C(m,E;\kappa)$ at each fixed pair $(E,\kappa)$. As evidenced by the identical distributions in Figs. S4b and S4g, as well as in Figs. S4c and S4f, the $\kappa$ dependency of $\mu$ respects the spectral symmetries of $\sigma(E(\theta)) = \sigma(E(-\theta))$ and $\sigma(E(\theta)) = -\sigma(E(\theta + \pi))$. Following the guideline in the previous work [1]—namely, that the best estimate of topological protection is given by the maximum local gap over $\kappa$ for the same local markers—we employ $\kappa = 10$ throughout all analyses to ensure consistency across $\theta$.

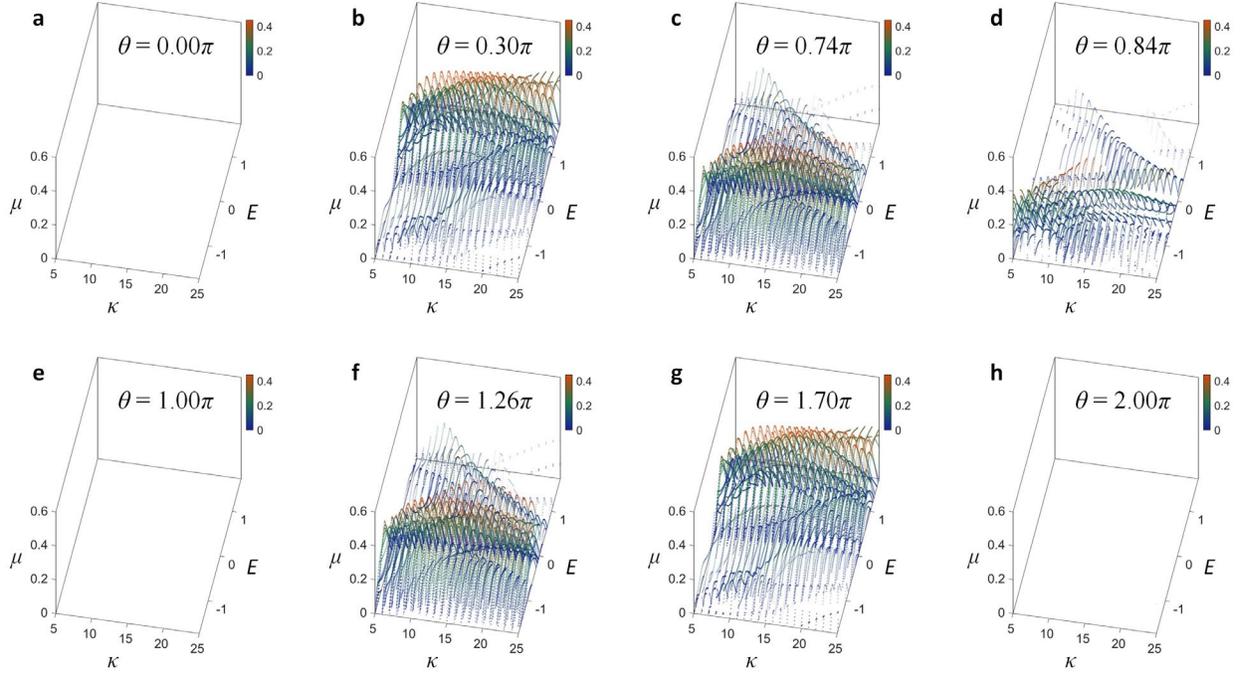

**Fig. S4. κ-dependence of local gaps.** Variations of $\mu$ with $\kappa$ and $E$ for different flux values: $\theta$ = 0.00 (**a**), $\theta$ = 0.30 (**b**), $\theta$ = 0.74 (**c**), $\theta$ = 0.84 (**d**), $\theta$ = 1.00 (**e**), $\theta$ = 1.26 (**f**), $\theta$ = 1.70 (**g**), and $\theta$ = 2.00 (**h**).

**Note S6. Dependence on perturbation strength**

Figure S5 shows the dependence of the nonreciprocal Bloch bands on the onsite perturbation strength $\delta_0$. The results indicate that the overall trends observed in Fig. 5 of the main text are well preserved, including the relative robustness near the large-local-gap region at $E = -0.886$, the degree-dependent spectral control (Figs. S5a–S5c), and the dominant role of peripheral nodes in driving the spectral response. We also note that perturbations applied to hub nodes do not induce substantial spectral control in all values of $\delta_0$ (Fig. S5d).

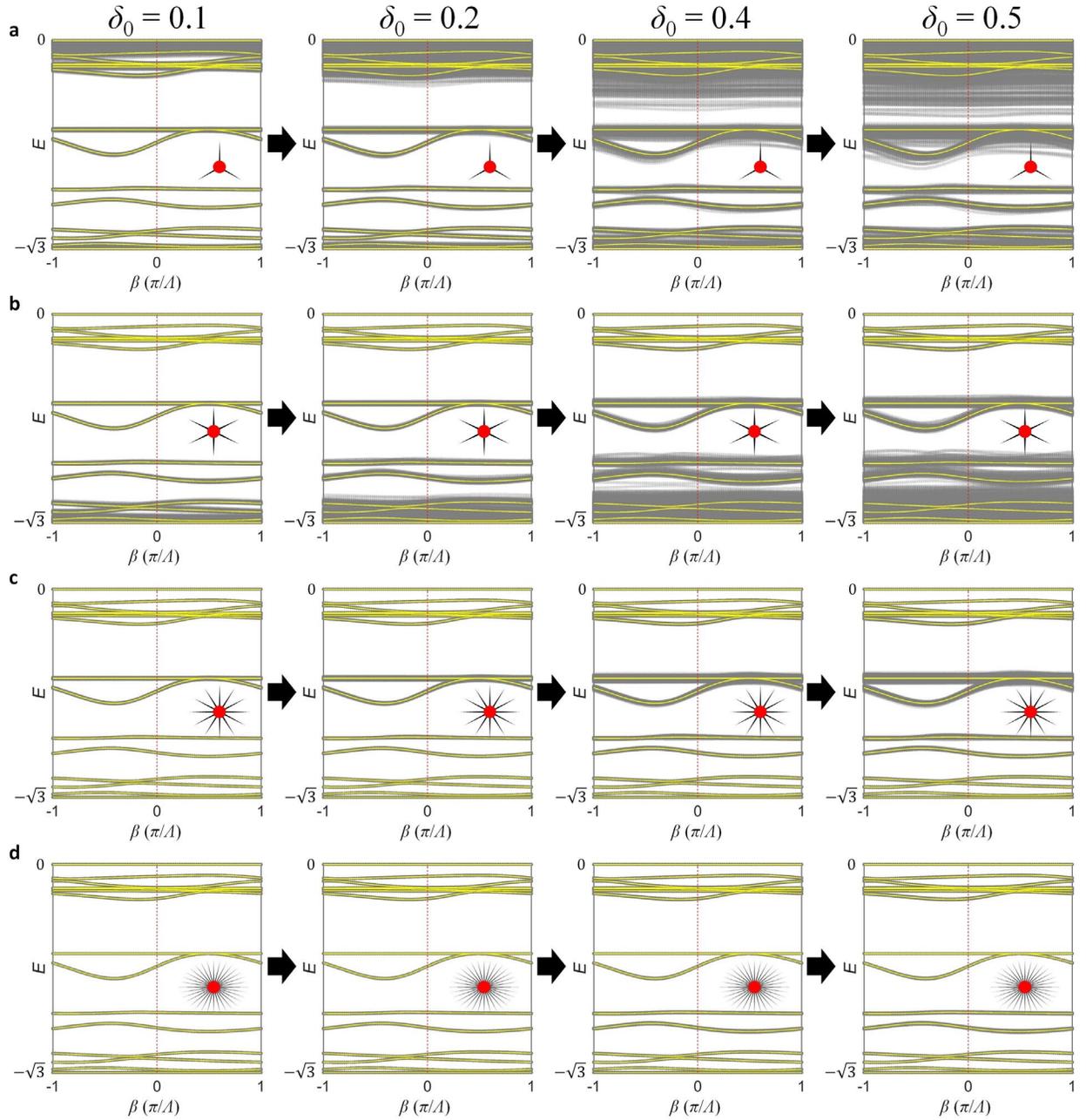

**Fig. S5. Dependence on the onsite perturbation strength. a-d,** Variations of the nonreciprocal Bloch bands $E(\beta)$ with $\delta_0$, for perturbations applied to nodes of degree 3 (**a**), 6 (**b**), 12 (**c**), and to hub nodes with degrees {24, 33, 48} (**d**). All the other parameters are the same as those in Fig. 5 of the main text.

**Note S7. Hub localization**

Figure S6 illustrates hub localization in the Apollonian network. To quantify wave localization, we use the inverse participation ratio (IPR):

$$\text{IPR} = \frac{\sum_m |\psi(m)|^4}{\left(\sum_m |\psi(m)|^2\right)^2}. \tag{S7}$$

Figure S6a shows the IPR of the Apollonian butterfly, revealing strong localization near the spectral boundaries, as indicated by the large IPR values. Together with the rapid convergence of the spectral geometry of the Apollonian butterfly near $E = 0$ as $g$ increases, this suggests that the spectral boundaries may be regarded as corresponding to the spatial boundaries of the finite Apollonian network.

The IPR analysis further reveals an intriguing form of hub localization in the eigenstate profile, clarifying which nodes correspond to the spatial boundary. Figure S6b shows the eigenstate profile of the one-dimensional Apollonian networks described in Fig. 5a of the main text, at $E = 6.75$ and $\beta = -0.5\pi/\Lambda$ for flux $\theta = 0.74\pi$. We note that, at this spectral boundary, the state is strongly localized on hub nodes in contrast to Figs. 5f-5h in the main text. This observation suggests that, in the strongly inhomogeneous Apollonian network, hub nodes correspond to the spatial boundaries that determine the finiteness of its geometry.

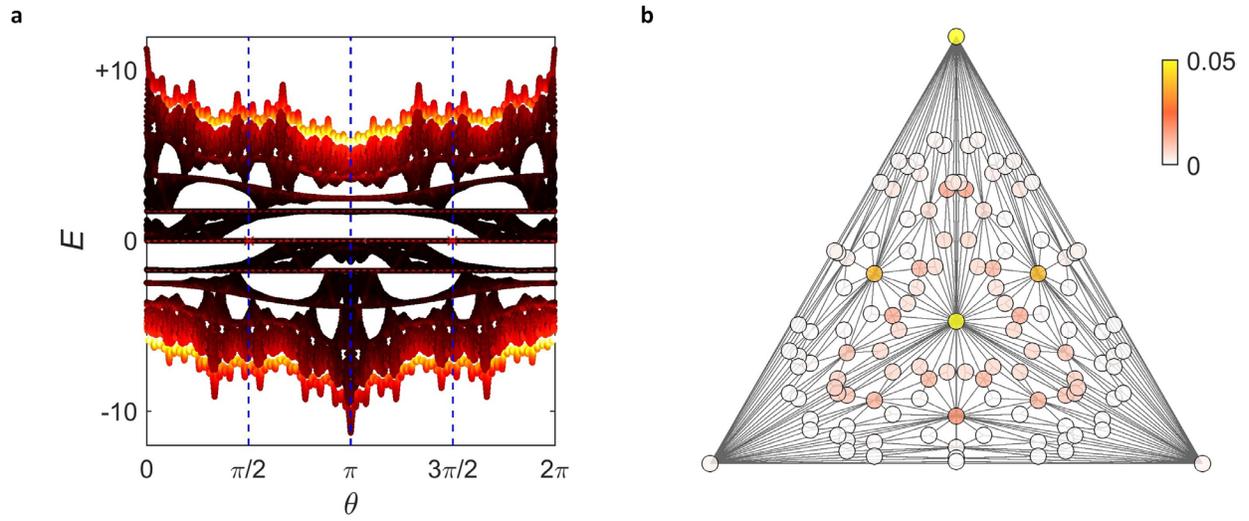

**Fig. S6. Localization at hub nodes. a,** IPR profile of the Apollonian network. **b,** The intensity of the wavefunctions $|\psi(m)|^2$ for the unperturbed state: $E = 6.75$ at $\beta = -0.5\pi/\Lambda$ and $\theta = 0.74\pi$.